\documentclass[runningheads]{llncs}
\usepackage{amsmath}
\usepackage{amssymb}
\usepackage{algorithm}
\usepackage{algorithmicx}
\usepackage{algpseudocode}
\usepackage{graphicx}
\graphicspath{{images/}}
\usepackage[section]{placeins}

\begin{document}
	\pagestyle{headings}
	\title{Rank Constrained Diffeomorphic Density Motion Estimation for Respiratory Correlated Computed Tomography\thanks{The final publication is available at \texttt{https://doi.org/10.1007/978-3-319-67675-3\_16}. This work was partially supported through research funding from the National Institute of Health (R01CA169102).}}
	\titlerunning{Rank Constrained Motion Estimation}
	\author{Markus D. Foote\inst{1} \and Pouya Sabouri\inst{2} \and Amit Sawant\inst{2} \and Sarang C. Joshi\inst{1}}
	\authorrunning{M. Foote et al.}
	\institute{Scientific Computing and Imaging Institute, Department of Biomedical Engineering, University of Utah\\%
		\email{foote@sci.utah.edu},%
		\and University of Maryland School of Medicine, Baltimore, Maryland}
	
	\maketitle
	\begin{abstract}
		Motion estimation of organs in a sequence of images is important in numerous medical imaging applications. The focus of this paper is the analysis of 4D Respiratory Correlated Computed Tomography (RCCT) Imaging.  It is hypothesized that the quasi-periodic breathing induced motion of organs in the thorax can be represented by deformations spanning a very low dimension subspace of the full infinite dimensional space of diffeomorphic transformations.   This paper presents a novel motion estimation algorithm that includes the constraint for low-rank motion between the different phases of the RCCT images. Low-rank deformation solutions are necessary for the efficient statistical analysis and improved treatment planning and delivery. Although the application focus of this paper is RCCT the algorithm is quite general and applicable to various motion estimation problems in medical imaging.
	\end{abstract}

	\section{Introduction}

\begin{figure}
	\centering
	\includegraphics[width=0.45\linewidth]{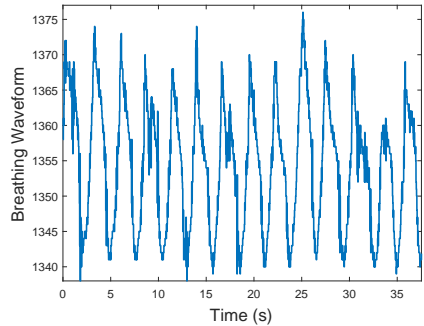}
	\caption{Breathing waveform of a RCCT subject. Variation in breathing intensity, rate, and pattern is apparent between respiratory cycles.}
	\label{fig:breathingwaveform}
\end{figure}

In this paper we consider the image registration problem for a set of images acquired over the breathing cycle by Respiratory Correlated Computed Tomography (RCCT). This problem has widespread medical applications, in particular 4D radiation therapy for lung cancer patients which considers lung deformations during treatment planning and delivery. Fundamental to the application of 4D motion modeling to improve radiation treatment planning and delivery is the statistical analysis of organ motion which can vary significantly from one breathing cycle to another \cite{Li2014}. Shown in Fig. \ref{fig:breathingwaveform} is a sample breathing trace captured by an abdominal belt in lung cancer radiation treatment patient. This cycle-to-cycle variability has recently been accounted for by live surface tracking methods in conjunction with Principal Component Analysis (PCA) of the deformation fields to develop a low dimensional representation of the motion (usually two). The use of Principal Component Analysis (PCA) to draw statistical relations between surface tracking data and RCCT is inherently lossy due to truncation of deformation fields to the few largest principal components \cite{Li2014,Sabouri2017}.

We extend pairwise weighted density matching first developed by Rottman et al. \cite{Rottman2015} for application to statistical analysis of the breathing cycle by incorporating a direct constraint on the rank of the estimated deformations and by considering an entire image series in single optimization problem. This method allows for the preservation of more descriptive deformations in downstream statistical processing that is dependent upon the rank of the deformation fields. Physiologically, the basis of density matching provides for tissue expansion and compression to occur within the lung while the low-rank optimization relates motion between all images in the series to describe the basic inhale-exhale breathing process very well, along with respiratory hysteresis.

Although the rank constraint introduced in this paper is applicable to any image registration algorithm, we focus on the Diffeomorphic Density Matching framework.
Density matching has previously been show to be very effective in pairwise RCCT image registration \cite{Rottman2015}. Considering the image volumes as densities provides the mathematical foundation to consider conservation of mass between images. Density action of the deformation on the image provides a mechanism through which compression of tissue results in an increased reported density by the deformed CT image, or vice-versa with tissue expansion \cite{Bauer2015}. This mathematical foundation also provides an efficient method for diffeomorphic registration, as integration of geodesic equations is avoided (contrary to methods like LDDMM \cite{Beg2005}).

	\section{Low Rank Motion Estimation}

Our problem extends the diffeomorphic density matching problem \cite{Rottman2016} to find a set of diffeomorphic transformations between one base image and a set of related images which exist in a low-rank subspace of the space of diffeomorphisms, $\text{Diff}(\Omega)$. 

Measuring the rank of the set of deformations is accomplished by the surrogate nuclear norm of the deformation matrix \cite{Recht2007}. Formal rank of the matrix, the number of non-zero eigenvalues, is avoided due to the non-smooth nature of the rank function. Instead, the nuclear norm serves as a convex surrogate function. The nuclear norm for a matrix $X$ is defined as 

\begin{equation}
\left\| X \right\|_* = \text{trace}\left(\sqrt{X^{*}X}\right) = \sum_{i}^{\min\{m,n\}} \sigma_i\left(X\right) %
\end{equation}
$\sigma_i$ is the $i$-th singular value of the $m\times n$ matrix $X$. Note that the singular values $\sigma_i$ are positive. We interpret each vectorized deformation field as a row of this matrix,
\begin{equation}
X=\left[\begin{matrix}
\varphi_1^{-1}(x) -x\\
\varphi_2^{-1}(x) -x\\
\vdots\\
\varphi_{N-1}^{-1}(x) - x

\end{matrix} \right]= \{\varphi_i^{-1}(x) - x\}
\end{equation}
where $\varphi_i^{-1}$ is the inverse of the deformation from the $i$-th image in the image series to a selected reference image. We can thus define the nuclear norm for deformations between $N$ images as 
\begin{equation}
\left\| X \right\|_* = \sum_{i}^{N-1} \sigma_i\left(X\right)
\end{equation}
as there are $N-1$ deformations between $N$ images, giving only $N-1$ singular values. The nuclear norm measure on this grouped deformation matrix effectively constrains the rank of the deformation set because of the summation of the singular values. 

The rank minimization builds upon the density matching framework, summarized here for completeness \cite{Bauer2015,Rottman2015}. A density or volume form $I\, dx$ is acted upon by a diffeomorphism $\varphi$ to compensate for changes of the density by the deformation:
\begin{equation}
\left(\varphi, I\, dx\right)\mapsto \varphi_*\left(I\, dx\right) = \left(\varphi^{-1}\right)^*\left(I\, dx\right)=\left(|D\varphi^{-1}|I\circ\varphi^{-1}\right)dx
\end{equation}
where $|D\varphi^{-1}|$ denotes the Jacobian determinant of $\varphi^{-1}$. The Riemannian geometry of the group of diffeomorphisms with a suitable Sobolev $H^1$ metric is linked to the Riemannian geometry of densities with the Fisher-Rao metric \cite{Bauer2015,Khesin2013,Modin2015}. The Fisher-Rao metric is used due to the property that it is invariant to the action of diffeomorphisms:
\begin{equation}
	d^2_F\left(I_0\, dx, I_1\, dx\right) = \int_{\Omega}\left(\sqrt{I_0}-\sqrt{I_1}\right)^2 dx \;.
\end{equation}

The linkage between a suitable Sobolev $H^1$ metric and the Fisher-Rao metric allows for evaluation of the distance in the space of diffeomorphisms in closed form. The Fisher-Rao metric and an incompressibility measure can then be used to match an image pair by minimizing the energy functional:
\begin{equation}
E\left(\varphi\right) = \int_{\Omega}\left(\sqrt{\left|D\varphi^{-1}\right|I_1\circ\varphi^{-1}}- \sqrt{I_0} \right)^2 dx \; + 
\int_{\Omega} \left(\sqrt{\left|D\varphi^{-1}\right|}-1\right)^2 f \,dx \;.
\end{equation}
The first term here penalizes dissimilarity between the two densities. The second term penalizes deviations from a volume-preserving deformation. The penalty function $f$ acts as weighting of the volume-preserving measure. A change of volume is penalized more (or less) where $f$ is large (or small). 

This problem has been solved by taking the Sobolev gradient of this energy functional and performing Euler integration of the gradient flow \cite{Rottman2015}:
\begin{multline}
\delta E = -\Delta^{-1} \left(-\nabla\left(f\circ\varphi^{-1}\left(1-\sqrt{|D\varphi^{-1}|}\right)\right)\right.\\\left.-\sqrt{|D\varphi^{-1}|I_1\circ\varphi^{-1}}\nabla\left(\sqrt{I_0}\right)+\nabla\left(\sqrt{|D\varphi^{-1}|I_1\circ\varphi^{-1}}\right)\sqrt{I_0}\right)
\end{multline}
\begin{equation}
\varphi^{-1}_{j+1}\left(x\right)=\varphi_j^{-1}\left(x+ \epsilon\delta E\right)
\end{equation}

We approach the rank constrained density matching problem by including the nuclear norm measure of the deformation fields matrix in the minimization problem and extending pairwise matching to the collective matching of a group of images to the reference image. We therefore seek to solve the following:
\begin{equation}
\begin{aligned}
&\underset{\{\varphi_i^{-1}\}}{\min} 
& &\sum_{i}^{N-1} \int_{\Omega}\left(\sqrt{\left|D\varphi_i^{-1}\right|I_i\circ\varphi_i^{-1}}- \sqrt{I_0} \right)^2 dx \; + 
 \int_{\Omega} \left(\sqrt{\left|D\varphi_i^{-1}\right|}-1\right)^2 f \,dx \\
& & & s.t. \quad  \left\| \{\varphi_i^{-1}(x) - x\} \right\|_* < k 
\end{aligned} \label{eq:problem}
\end{equation}
where $I_0$ is a chosen base or reference image and $I_i$ are the other $N-1$ images in the series. We re-frame the rank constraint as a Lagrange multiplier to include the nuclear norm rank measure as a penalty function.  This formulation allows us to directly apply the rank minimization strategies such as the iterative shrinkage-thresholding algorithm (ISTA) outlined by Cai et al. \cite{Cai2010}. 
Our problem can thus be written as the minimization of the following energy functional:
\begin{multline}
E(\left\{\varphi_i\right\}) = \sum_{i}^{N-1}\left[ \int_{\Omega}\left(\sqrt{\left|D\varphi_i^{-1}\right|I_i\circ\varphi_i^{-1}}- \sqrt{I_0} \right)^2 dx \;  \right. \\
\left. +\int_{\Omega} \left(\sqrt{\left|D\varphi_i^{-1}\right|}-1\right)^2 f \,dx \right]  + \alpha \sum_{i}^{N-1} \sigma_i\left(\{\varphi_i^{-1}(x) - x\}\right)  \;.
\label{eq:energy}
\end{multline}

	\section{Singular Value Thresholding and Implementation}

In this section we describe in detail our implementation of the solution to (\ref{eq:problem}) by the ISTA algorithm, with special consideration for efficient acceleration by GPGPU programming through the PyCA software package \cite{pyca}.

This problem seeks to minimize the singular values of the deformations, so we perform ISTA \cite{Cai2010} on the singular value decomposition of the ideal $H^1$ gradient of the diffeomorphisms.
The shrinkage-thresholding algorithm is employed by the shrinkage operator \cite{Cai2010}:
\begin{equation}
\mathcal{D}_\tau \left(\Sigma\right) = \text{diag}\left(\{\sigma_i-\tau\}_+\right)
\end{equation}
where the singular value decomposition is noted as $X=U\Sigma V^*$, thus the shrinkage acts only on the singular values, and $t_+=\max(0,t)$.

The solution to (\ref{eq:problem}) can therefore be found through an ISTA approach by first finding an optimal update for the density matching problem of each image pair, then performing the shrinkage operation on singular values of the updated fields, and finally replacing the deformations with reconstructions by SVD of the shrunken singular values. 
In our implementation, we choose to perform SVD on the deformation gram matrix %
$XX^*$, as our GPU image processing library lacks an SVD algorithm. This allows for accelerated computation of the gram matrix instead of an accelerated SVD, and only a small penalty for performing SVD on a small $9\times 9$ matrix on the host CPU. Combining the solution to a single density matching problem with our singular value thresholding algorithm gives the algorithm: 

\begin{algorithm}
	\caption{GPU Accelerated Algorithm \label{alg:final}}
	\begin{algorithmic}
		\State Choose step size $\epsilon >0$
		\State Choose rank weighting parameter $\alpha >0$
		\State Set $\varphi_{i}^{-1} = \text{id}$
		\State Set $\left| D\varphi^{-1}_i\right| = 1$
		\For{$iter$ = 1 .. NumIter}
		\For{i = 1 .. $N-1$}
		\State Compute $\varphi_{*i}I_i = I_i\circ\varphi_i$
		\State Compute $u=-\nabla \big(f\circ\varphi_i^{-1}(1-\sqrt{|D\varphi_i^{-1}|})\big) - \sqrt{\varphi_{*i}I_i}\nabla\sqrt{I_0} + \nabla(\sqrt{\varphi_{*i}I_i})\sqrt{I_0}$
		\State Compute $v=-\Delta^{-1}(u)$
		\State Update $\varphi_i^{-1} \to \varphi_i^{-1}(x + \epsilon v)$
		\EndFor
		\State Compute $\vec{K} = XX^*$%
		\State Compute $\vec{U} \vec{\Sigma} \vec{V^*} = \vec{K}$ on host CPU
		\State Compute $\vec{W} = \vec{U}\mathcal{D}_{\epsilon\alpha}(\vec{\Sigma})$ on host CPU
		\State Update $\{\varphi^{-1}_i\} \to \vec{W}X + x$
		\State Compute $\left| D\varphi^{-1}_i\right|$
		\EndFor
	\end{algorithmic}
\end{algorithm}

We further accelerate the above algorithm by implementing a multi-scale approach. Rather than use the full resolution data from initialization, the algorithm is instead initialized at a lower resolution with down-sampled data. After convergence at the lower resolution, a lower down-sampling factor is selected, resulting in a resolution closer to full resolution. At each scale level change the current deformation field estimates are up-sampled to the new scale and the data is again down-sampled from the original, full resolution images. The final scale level is at the same resolution of the original data.

This multi-scale approach requires two special considerations for tracking the energy being minimized. First, as the volume of a voxel is not constant, the penalties from a voxel must be scaled by the current voxel volume. In other words, the energy must be considered volumetrically, not simply as a data grid. Second, the gram matrix $K$ must be divided by the number of voxels, as a scale change results in the summation over millions more voxels of the deformation fields which would otherwise greatly increase the singular values. Inclusion of these two scale-dependent factors allows the total energy of (\ref{eq:energy}) to be tracked over the multiple scale levels without massive increases when the scale level is changed.
	\section{Application to Respiratory 4DCT Phase Registration}

A RCCT of a radiotherapy patient was acquired at University of Maryland and provided as 10 respiratory phase-binned images. The full exhale image was chosen as the reference image for the registration problem. Image intensities were modified with an exponential function as in \cite{Rottman2016} to transform the intensity such that the volume exhibits conservation of mass.
The final deformations were computed at the resolution of the original 3D volume ($320 \times 256 \times 144$); all the figures show the same middle sagittal, coronal, and axial slices of the volume. 

For the compressibility penalty $f$, we used a soft thresholding of the intensity values of the base image using the logistic function. High intensity regions were penalized with $5\sigma$ as dense, incompressible tissue, and vice-versa for low intensity regions ($0.2\sigma$). The incompressibility parameter, $\sigma$, was set at 0.01 for all runs. The algorithm was implemented on a single Nvidia GTX Titan X GPU, which runs 1000 iterations of the full-resolution volume in approximately 17 minutes for all 10 images. Lower scales of the multi-scale optimization run significantly faster, mainly due to the $O(n^2)$ complexity of calculating the gram matrix.

\begin{figure}
	\centering
	\begin{tabular}{ccc}
		Full Inhale & Full Exhale & Inhale Deformed 
		\\
		\includegraphics[width=0.33\linewidth, trim={120px 40px 120px 40px}, clip]{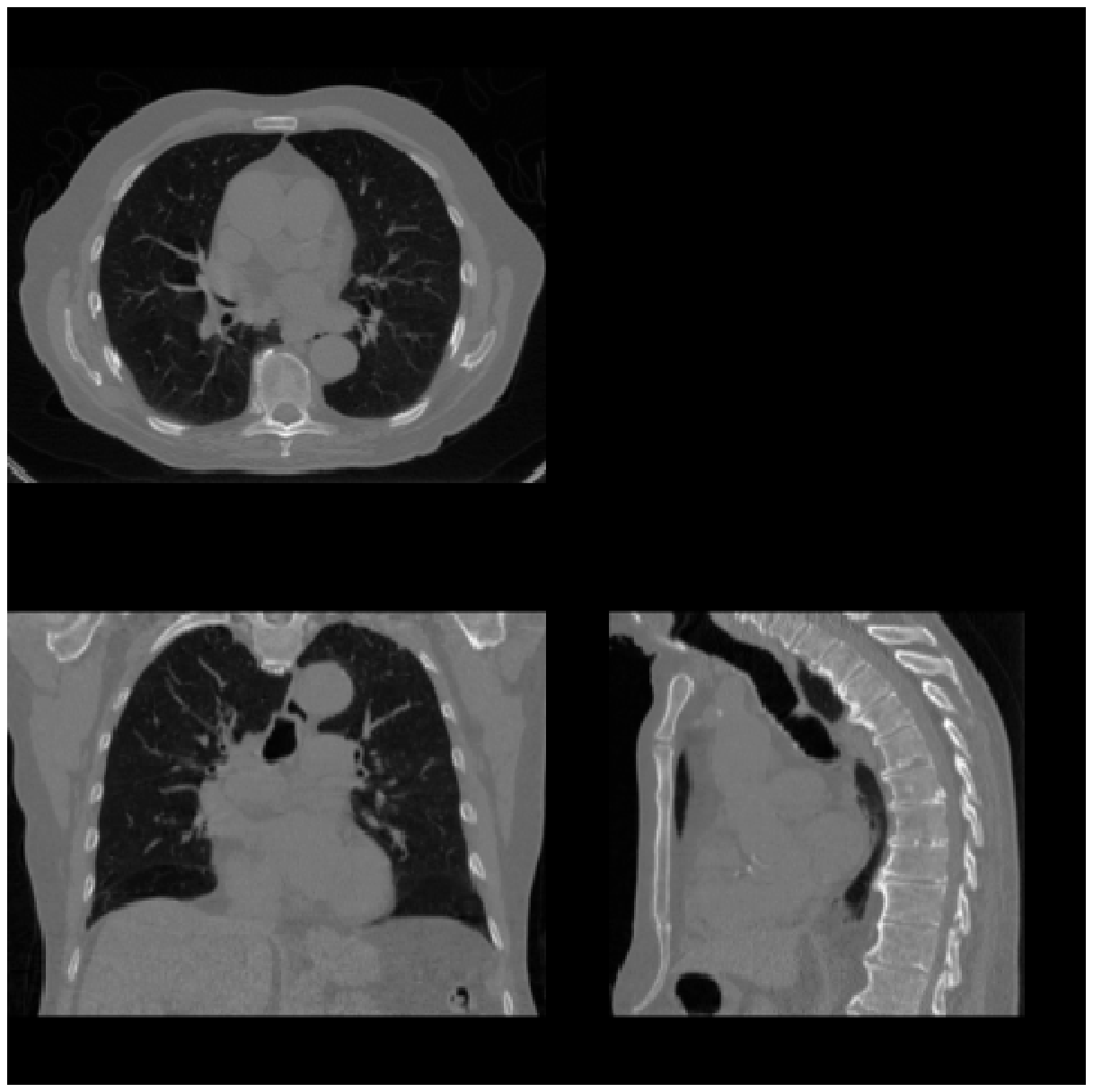}
		&
		\includegraphics[width=0.33\linewidth, trim={120px 40px 120px 40px}, clip]{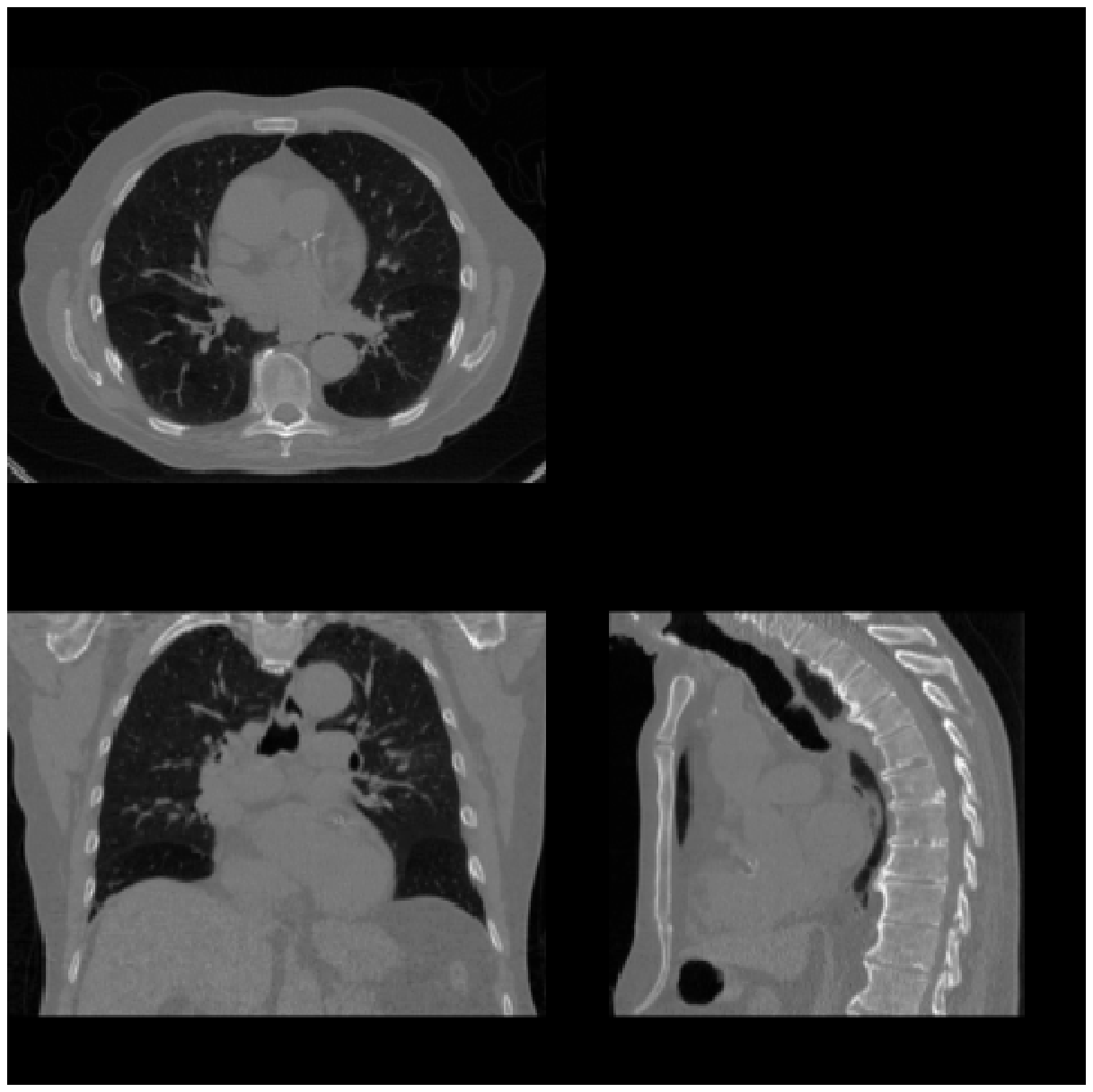}
		&
		\includegraphics[width=0.33\linewidth, trim={120px 40px 120px 40px}, clip]{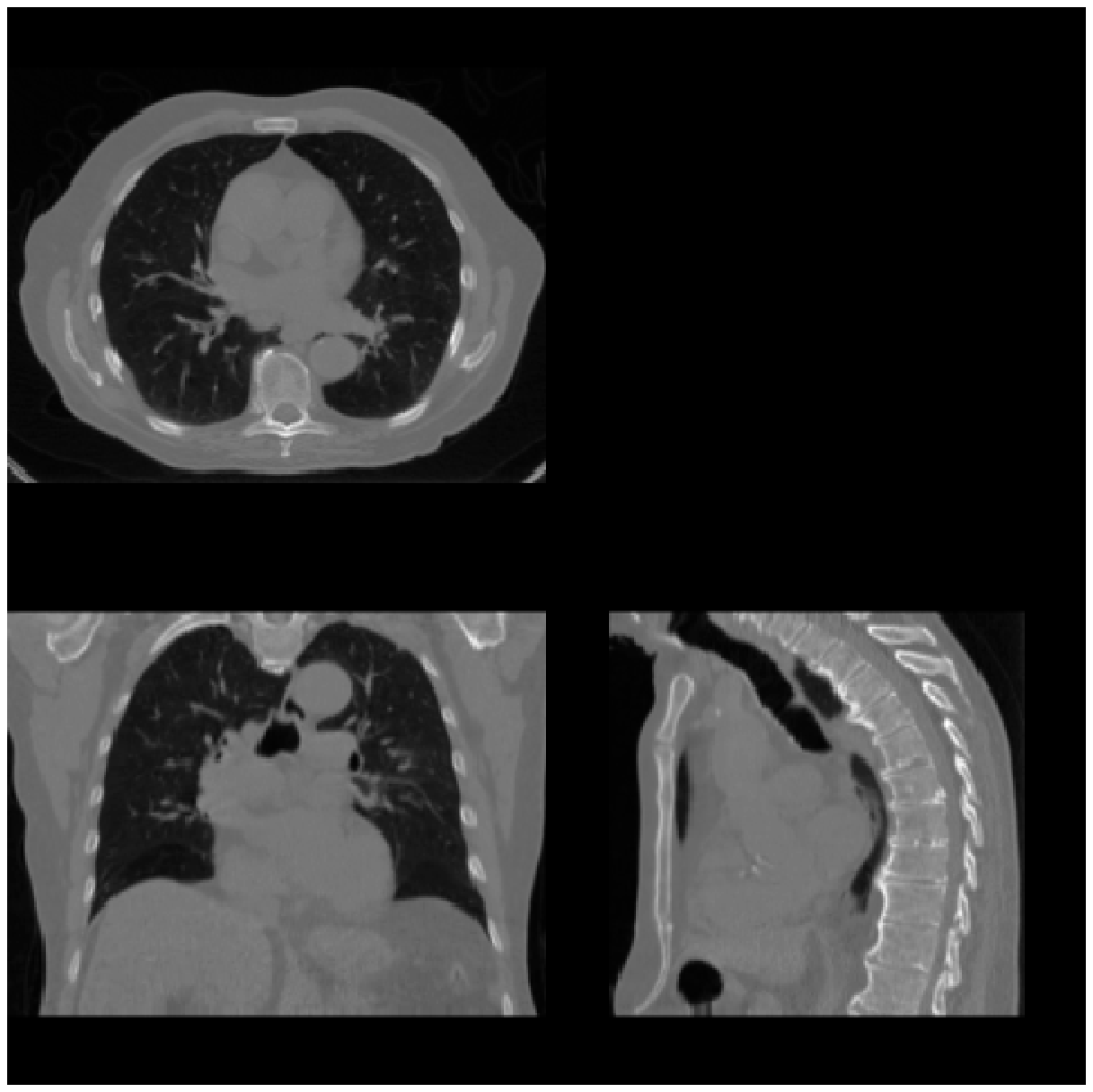}
		\\
		Jacobian Determinant & Energy & Penalty Function 
		\\
		\includegraphics[width=0.33\linewidth, trim={75px 35px 75px 35px}, clip]{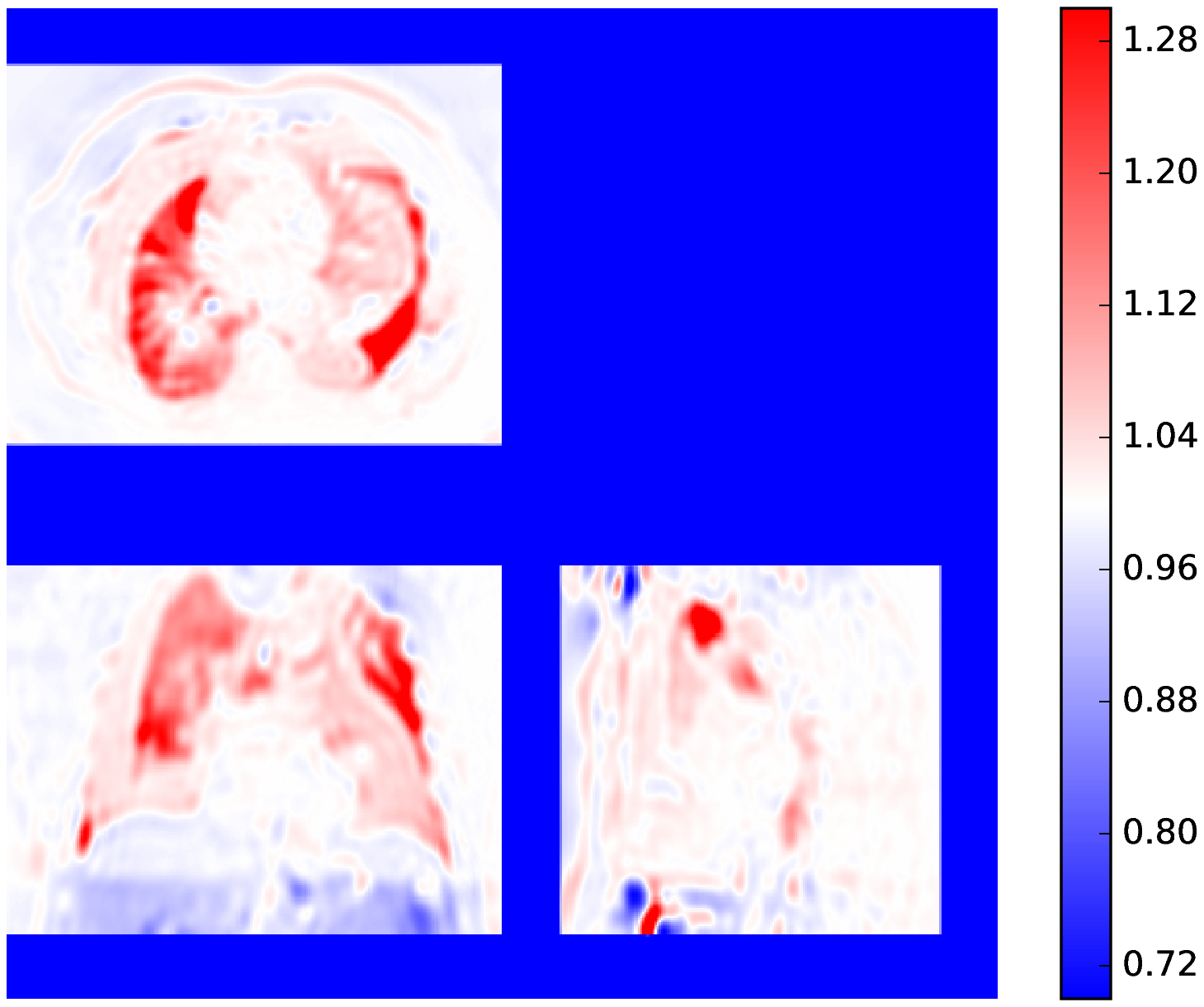}
		&
		\includegraphics[width=0.33\linewidth, trim={5px 15px 25px 35px}, clip]{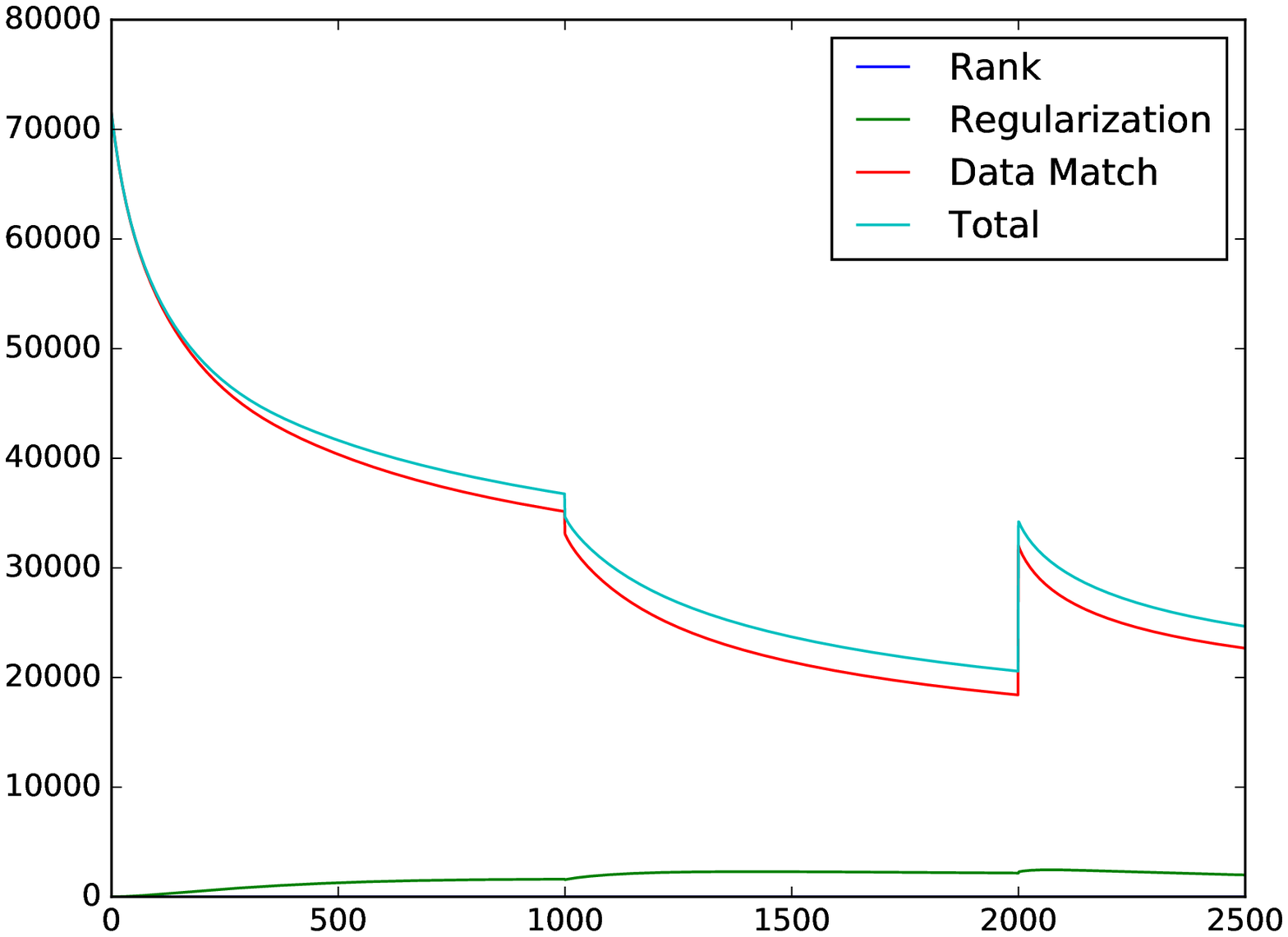}
		&
		\includegraphics[width=0.33\linewidth, trim={75px 35px 85px 35px}, clip]{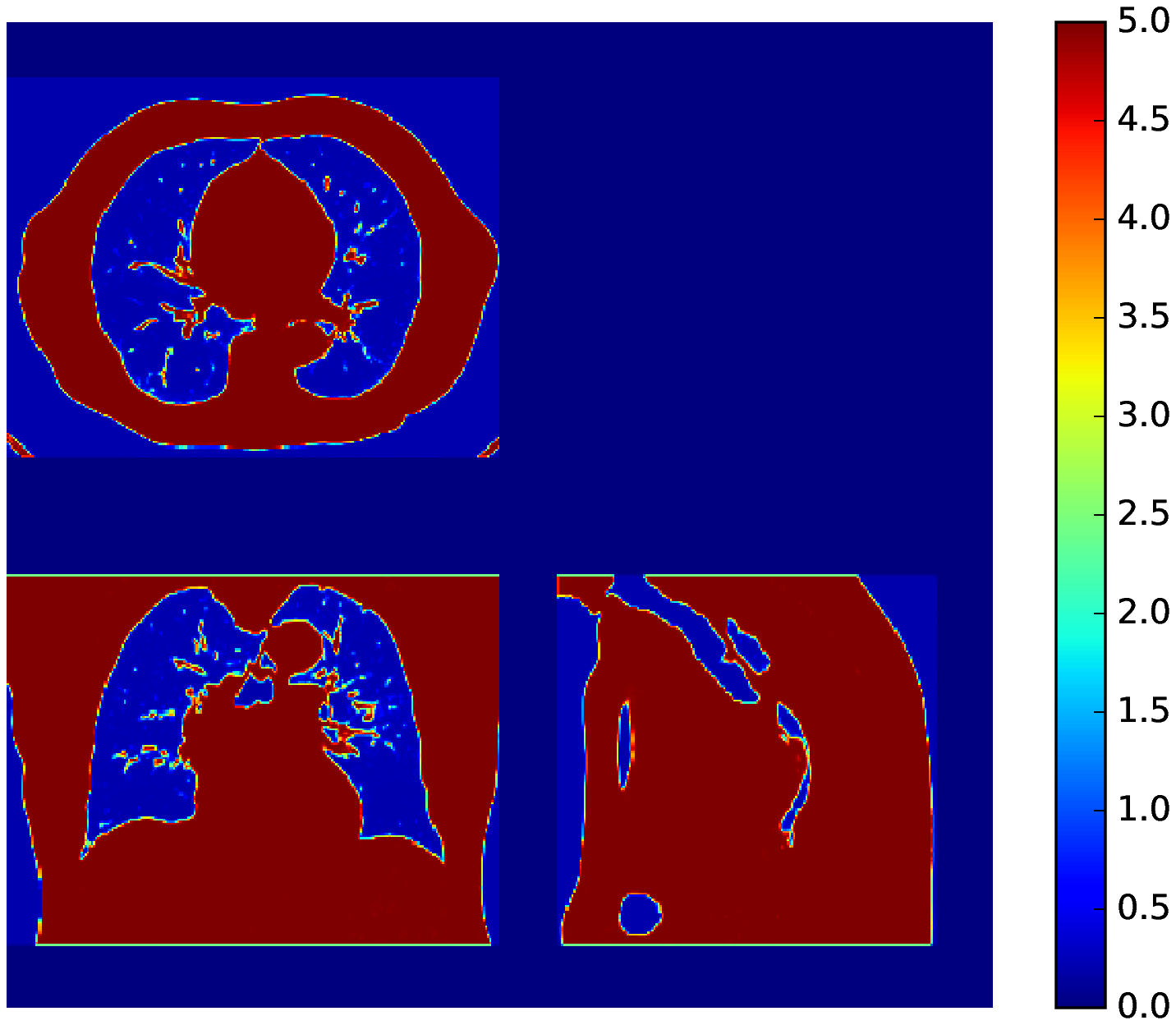}
	\end{tabular}
	\caption{Registration results for $\alpha=0.01$. Top row: Full inhale image, full exhale image, and registered inhale image to exhale. Bottom row: Jacobian determinant of the deformation to full exhale, energy plot, and penalty function for density matching algorithm. Note the energy plot shows three scale levels of a multi-scale run; the increase at 2000 is due to the first two scale levels having a blurring applied in the down-sampling procedure which removes noise in the data.}
	\label{fig:regresults}
\end{figure}

\begin{figure}
	\centering
	\begin{tabular}{ccc}
		GTV DICE & PTV DICE & Lung Volume DICE 
		\\
		\includegraphics[width=0.33\linewidth]{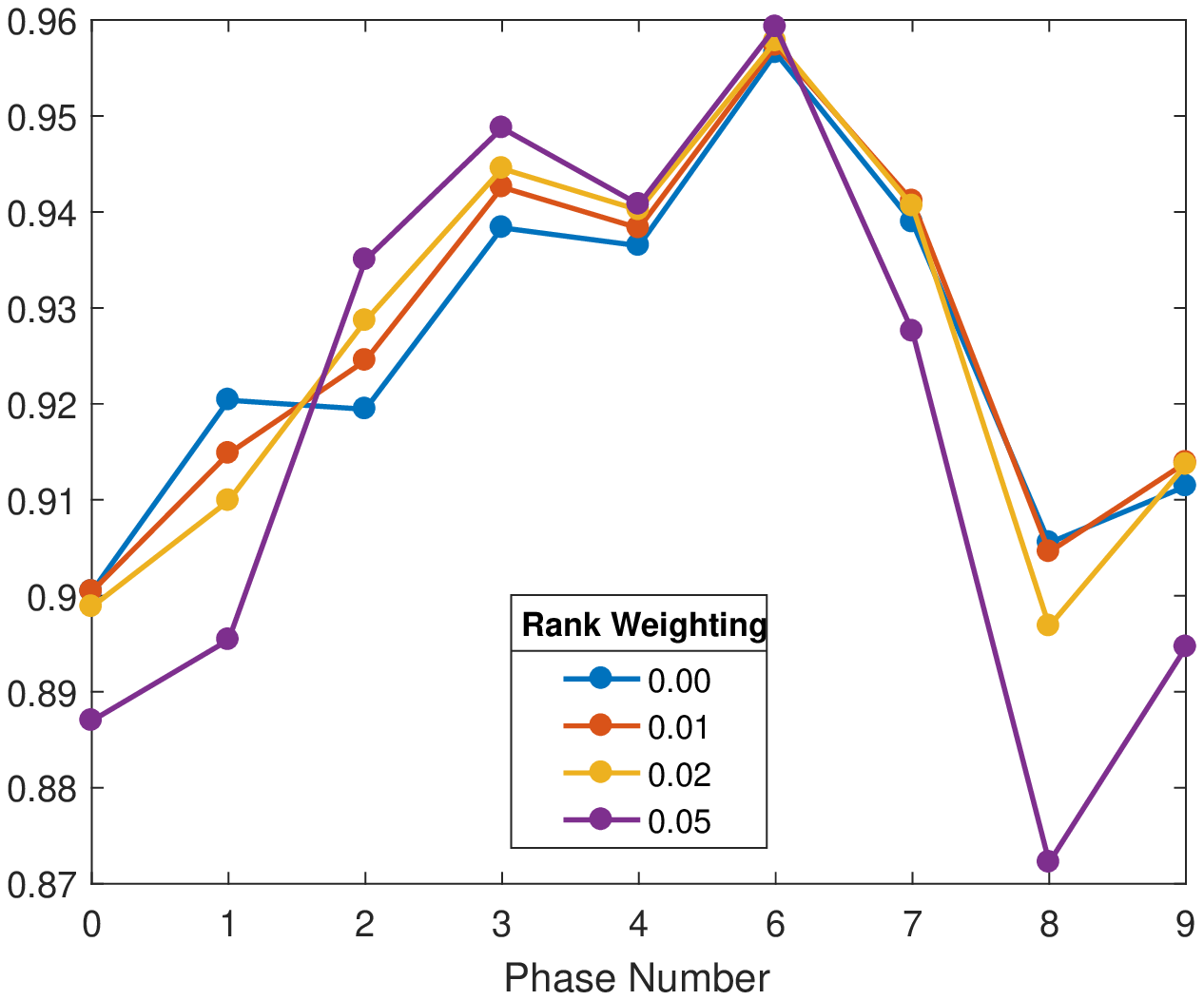}
		&
		\includegraphics[width=0.33\linewidth]{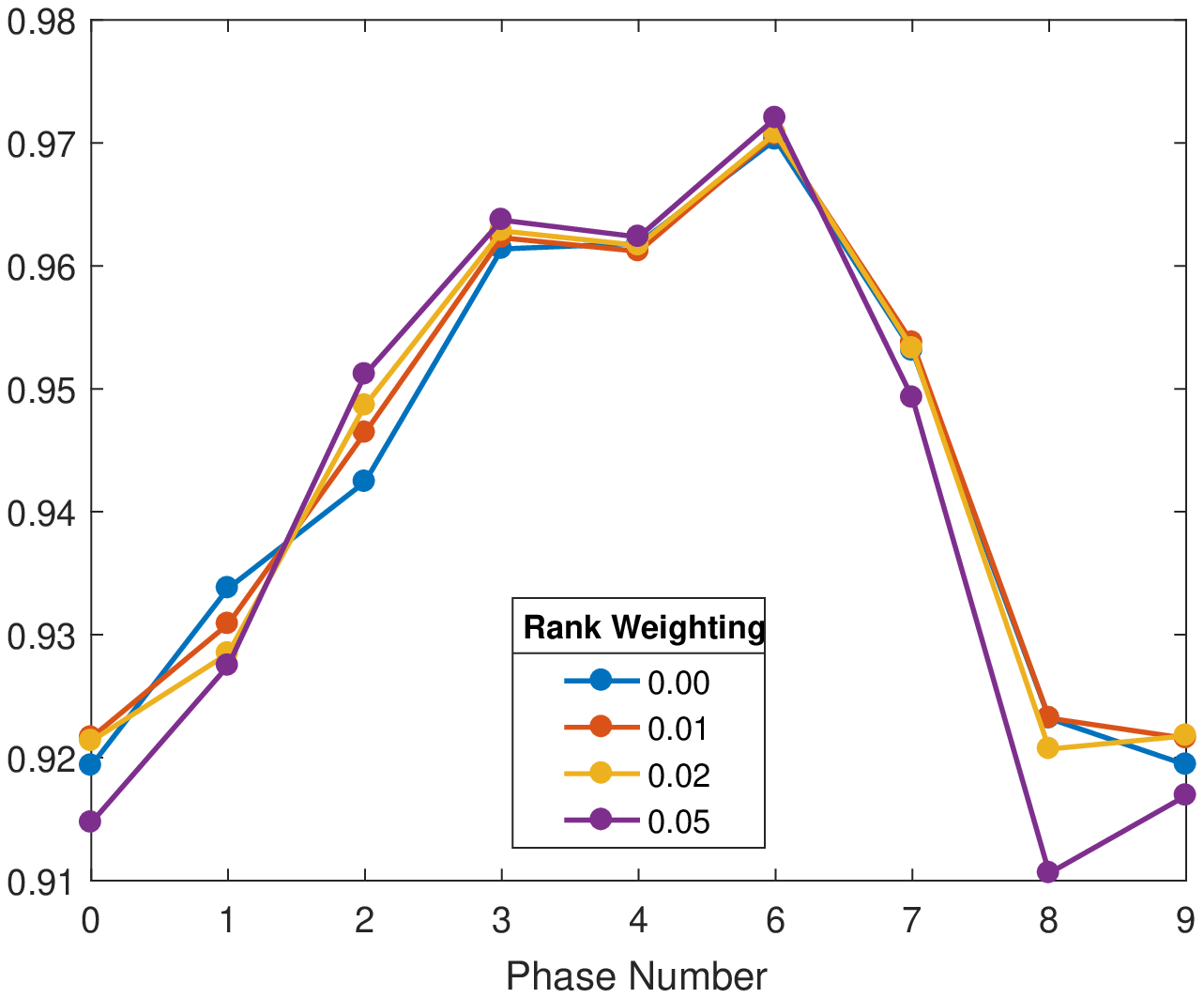}
		&
		\includegraphics[width=0.33\linewidth]{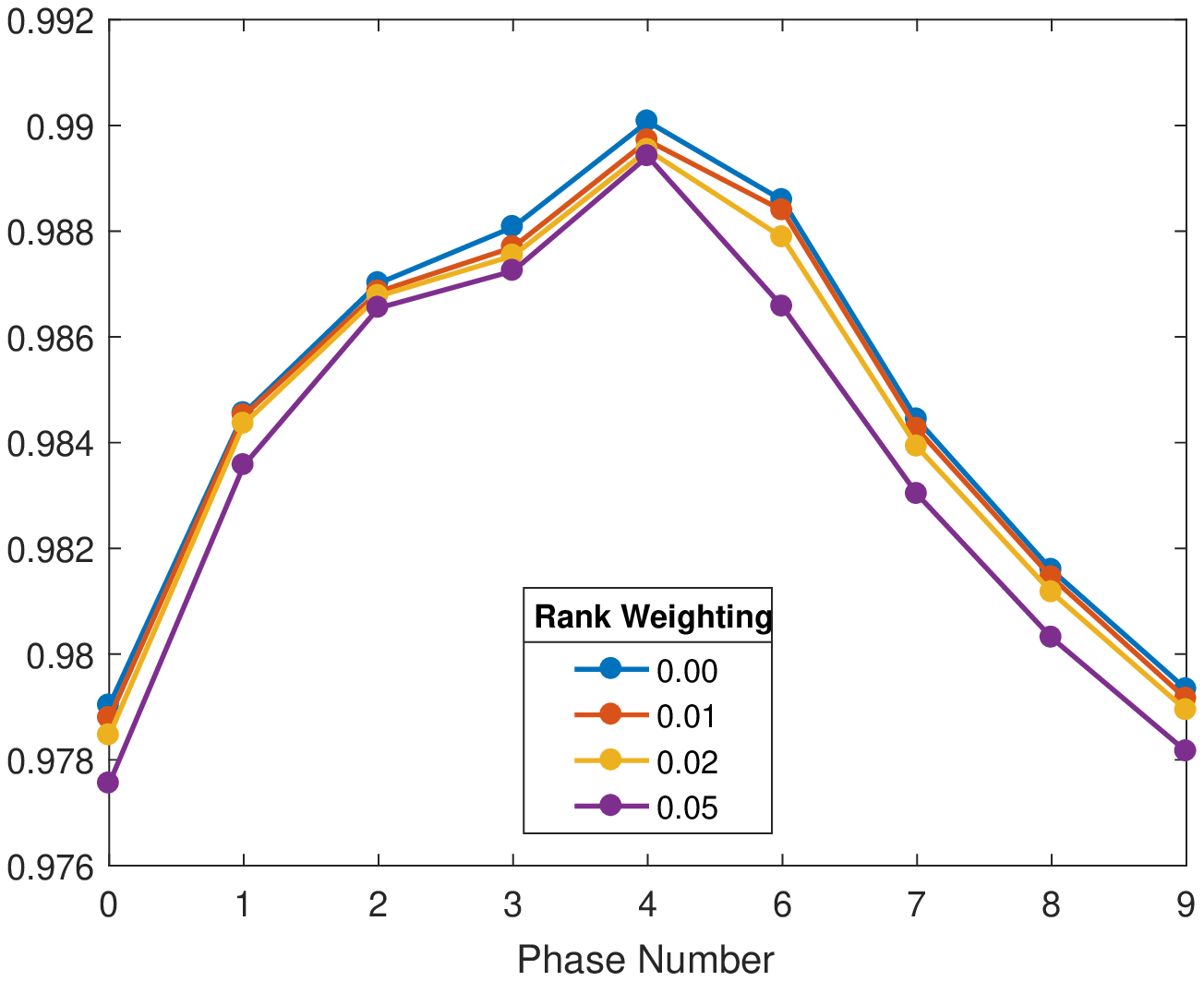}
	\end{tabular}
	\caption{DICE coefficients for registration results with various rank weightings, $\alpha$. $\quad$ GTV - Gross Tumor Volume, PTV - Planned Treatment Volume }
	\label{fig:dice}
\end{figure}

Deformations were calculated from each of the 9 other images with rank weighting parameter $\alpha$ of 0, 0.01, 0.02, and 0.05. Figure \ref{fig:regresults} shows the result of registration for one of the nine pairings, full inhale to full exhale. These deformations have geometric accuracy similar to that attained by the density matching without a rank constraint, as measured by the DICE coefficient between reference and deformed volumes (Fig. \ref{fig:dice}). 


Deformations resulting from the rank constrained algorithm are physiologically relevant, as with previous density matched results, because compression occurs predominantly within the lung tissue. Additionally, the confinement to a low-rank subspace of deformations requires relation to develop between the deformation fields, resulting in linkage of the generally reverse relation between inhalation and exhalation. This added rank constraint results in even better geometric accuracy of some motion estimates as measured by the DICE coefficients.

\begin{figure}
	\centering
	\includegraphics[width=0.45\linewidth]{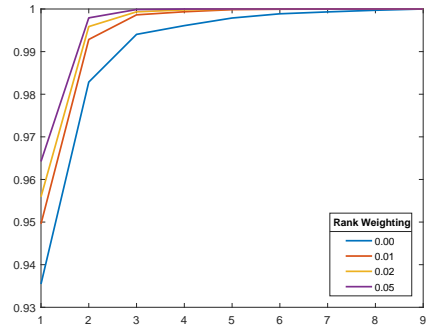}
	\caption{Normalized cumulative sum of singular values for registration results with various rank weightings,  $\alpha$. This effectively shows the percentage of the deformation fields that are explained by a number of principal components. Increased rank weighting produces deformations well-described by fewer principal components. }
	\label{fig:svd}
\end{figure}

\begin{figure}
	\centering
	\includegraphics[width=0.45\linewidth]{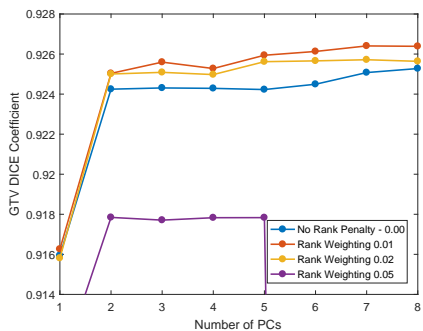}
	\caption{DICE Coefficients averaged across phases after reconstruction of deformations using a variable number of principal components. Including an appropriate rank constraint in the minimization results in more accurate deformation fields after a statistical truncation of the lower principal components.}
	\label{fig:truncateddice}
\end{figure}

Increased weighting of the rank term in the minimization problem produces sets of deformations that can be explained by fewer principal components (Fig. \ref{fig:svd}). The resulting deformations preserve geometric accuracy better when using PCA to truncate the deformation fields to the largest principal components. The average GTV DICE coefficient across all phases is shown for each number of principal components included in the reconstructed deformation field for a motion estimate performed with and without rank constraint in Fig. \ref{fig:truncateddice}. Further increase of the rank weighting (such as 0.05), while effective at minimizing rank, causes significant loss in the anatomical accuracy of the deformation estimates (Fig. \ref{fig:dice}).

	\section{Discussion}

In this paper, we have shown that including rank minimization in the motion estimation problem improves deformation accuracy in later statistical analysis while improving anatomic accuracy. We implemented ISTA to minimize the rank of the deformations between a set of CT images throughout a breathing cycle. In particular, a rank weighting of 0.01 produces better overall geometric accuracy with a significant shift in the rank of the deformations which preserves the deformation accuracy through PCA treatment planning procedures. The geometric accuracy improvement may arise from increased physiologic relevance of the low-rank deformations matching well with the general reversal process of an inhale-exhale cycle, along with hysteresis in other components.
Substantial improvement in speed of our algorithm could be achieved by implementing a FISTA technique \cite{Beck2009}. Additional parallelization from upcoming multi-GPU systems  would provide a speedup with low complexity increase, as the density matching portion of the algorithm is completely independent between phases.

	\bibliography{mfca2017}

\begin{thebibliography}{10}
\providecommand{\url}[1]{\texttt{#1}}
\providecommand{\urlprefix}{URL }

\bibitem{Bauer2015}
Bauer, M., Joshi, S., Modin, K.: {Diffeomorphic Density Matching by Optimal
  Information Transport}. SIAM Journal on Imaging Sciences  8(3),  1718--1751
  (jan 2015),
  \url{http://arxiv.org/abs/1501.07635{\%}0Ahttp://dx.doi.org/10.1137/151006238
  http://epubs.siam.org/doi/10.1137/151006238}

\bibitem{Beck2009}
Beck, A., Teboulle, M.: {A Fast Iterative Shrinkage-Thresholding Algorithm}.
  Society for Industrial and Applied Mathematics Journal on Imaging Sciences
  2(1),  183--202 (2009)

\bibitem{Beg2005}
Beg, M.F., Miller, M.I., Trouv{\'{e}}, A., Younes, L.: {Computing Large
  Deformation Metric Mappings via Geodesic Flows of Diffeomorphisms}.
  International Journal of Computer Vision  61(2),  139--157 (feb 2005),
  \url{http://link.springer.com/10.1023/B:VISI.0000043755.93987.aa}

\bibitem{Cai2010}
Cai, J.F., Cand{\`{e}}s, E.J., Shen, Z.: {A Singular Value Thresholding
  Algorithm for Matrix Completion}. SIAM Journal on Optimization  20(4),
  1956--1982 (2010), \url{http://epubs.siam.org/doi/10.1137/080738970}

\bibitem{Khesin2013}
Khesin, B., Lenells, J., Misio{\l}ek, G., Preston, S.C.: {Geometry of
  Diffeomorphism Groups, Complete integrability and Geometric statistics}.
  Geometric and Functional Analysis  23(1),  334--366 (feb 2013),
  \url{http://link.springer.com/10.1007/s00039-013-0210-2}

\bibitem{Li2014}
Li, R., Lewis, J.H., Jia, X., Zhao, T., Liu, W., Wuenschel, S., Lamb, J., Yang,
  D., Low, D.A., Jiang, S.B.: {On a PCA-based lung motion model}. Physics in
  Medicine and Biology  56(18),  6009--6030 (sep 2011),
  \url{http://stacks.iop.org/0031-9155/56/i=18/a=015?key=crossref.2101ba1e0fc5d7788678ba73f94eef52}

\bibitem{Modin2015}
Modin, K.: {Generalized Hunter–Saxton Equations, Optimal Information
  Transport, and Factorization of Diffeomorphisms}. The Journal of Geometric
  Analysis  25(2),  1306--1334 (apr 2015),
  \url{http://link.springer.com/10.1007/s12220-014-9469-2}

\bibitem{pyca}
Preston, J., Hinkle, J., Singh, N., Rottman, C., Joshi, S.: {PyCA: Python for
  Computational Anatomy}, \url{https://bitbucket.org/scicompanat/pyca}

\bibitem{Recht2007}
Recht, B., Fazel, M., Parrilo, P.A.: {Guaranteed Minimum-Rank Solutions of
  Linear Matrix Equations via Nuclear Norm Minimization}. SIAM Review  52(3),
  471--501 (jan 2010), \url{http://epubs.siam.org/doi/10.1137/070697835}

\bibitem{Rottman2015}
Rottman, C., Bauer, M., Modin, K., Joshi, S.C.: {Weighted Diffeomorphic Density
  Matching with Applications to Thoracic Image Registration}. 5th MICCAI
  Workshop on Mathematical Foundations of Computational Anatomy (MFCA 2015) pp.
  1--12 (2015)

\bibitem{Rottman2016}
Rottman, C., Larson, B., Sabouri, P., Sawant, A., Joshi, S.: {Diffeomorphic
  Density Registration in Thoracic Computed Tomography}. In: Ourselin, S.,
  Joskowicz, L., Sabuncu, M.R., Unal, G., Wells, W. (eds.) Medical Image
  Computing and Computer-Assisted Intervention -- MICCAI 2016: 19th
  International Conference, Athens, Greece, October 17-21, 2016, Proceedings,
  Part III, Lecture Notes in Computer Science, vol. 9902, pp. 46--53. Springer
  International Publishing (2016),
  \url{http://link.springer.com/10.1007/978-3-319-46726-9{\_}6
  http://dx.doi.org/10.1007/978-3-319-46726-9{\_}6}

\bibitem{Sabouri2017}
Sabouri, P., Foote, M., Ranjbar, M., Tajdini, M., Mossahebi, S., Joshi, S.,
  Sawant, A.: {A Novel Method Using Surface Monitoring to Capture
  Breathing-Induced Cycle-To-Cycle Variations with 4DCT}. In: 59th Annual
  Meeting of The American Association of Physicists in Medicine. Denver, CO
  (2017),
  \url{http://www.aapm.org/meetings/2017AM/PRAbs.asp?mid=127{\&}aid=37742}

\end{thebibliography}
	
\end{document}